\documentclass[12pt]{article}
\usepackage{amsfonts,amssymb,latexsym,oldgerm,amscd}
\title{\textbf{Avoidance of Singularity and Global Non-Conservation of Energy in General Relativity}}
\date{}
\author{ Murli Manohar Verma \footnote{sunilmmv@yahoo.com, mverma0@ictp.it}\\
{\small Department  of  Physics}\\
{\small Lucknow  University,  Lucknow  226 007, India}\\
{\small and}\\
{\small The Abdus  Salam  International  Centre  for  Theoretical Physics}\\
{\small Trieste,  Italy}\\
}

\begin{document}
\maketitle


 \abstract

 {We show that the singularity in the General Theory of Relativity (GTR) is the expression of a non-Machian feature. It can be avoided with a scale-invariant dynamical theory, a property lacking in GTR. It is further argued that the global non-conservation of energy in GTR also results from the lack of scale-invariance and the field formulation presented by several authors can only resolve the problem in  part. Assuming the global energy conservation, we propose a negative energy density component with positive equation of state that can drive the late-time acceleration in the universe, while the positive component confines to smaller scales.}



\pagebreak

\section{Introduction}

Over the past several decades the problems pertaining to singularity and global non-conservation of energy-momentum in General Theory of Relativity (GTR) have been discussed in detail \cite{a1,b1}.  However,it is seen that in most of the approaches to avoid singularity in GTR, it is generally achieved by dropping the assumptions in the singularity theorems considering the total  energy density contributed by  various components in the universe and by making a specific choice of the metric \cite{c1,d1,e1}. This is obviously not required by the dynamical theory on its own, but is dictated by the self-imposed constraints of our understanding. Therefore, it is not surprising that this allows one to construct a number of singularity-free models without any special requirements made by the theory. However, we emphasize that(considering that the metric cannot be measured directly)such an approach is flawed in that the freedom from singularity must stem from the dynamical theory,here GTR, and not from any subjective dropping of the conditions on singularity theorems or the choice of the metric. Since field equations are obtained from the  GTR action integrated over all spatial volume bound between the  arbitrary time limits, it is not clear why an epoch of singularity, which is later revealed as such by its field equations, must then  be surgically removed from it. If singularity is a truly unwanted feature, the theory must self-impose that condition at the beginning in its structure.

In Section 2 of this paper, first we show that the singularity is a non-Machian \cite{f1} feature of GTR and second, that a more fundamental approach via conformal invariance is required to handle singularity. As an example, we discuss it in the framework of the strongest version of the Machian principle found in the inter-particle interaction based Quasi-Steady State Cosmology(QSSC) which is both conformally invariant and singularity-free \cite{g1,g2,g3}. We attempt to highlight the issues of Machian connections and conformal invariance as a ground-work for  any flawless theory of gravity.
Section 3 focuses on another fundamental difficulty in defining energy in GTR that continues to intrigue many researchers for a long time now\cite{h1}. Some authors have proposed a  pseudotensor for gravity\cite{i1} or put some assumptions on the metric\cite{j1,k1} or proposed field formulation instead of a geometrical one\cite{l1,l2,l3}. We discuss this aspect of energy-momentum conservation in inescapable presence of gravity and its possible connection with the previous problem of singularity. We stress that the non-localizability of energy in GTR is related to a fact of inconsistency  between the equivalence principle and the conformal invariance.  Thus, by  zooming in to the infinitesimally small regions around a spacetime point, the validity of conformal transformation breaks and scale invariance loses meaning. This leads to the conclusion that non-localizability  of gravitational energy appears to be a  consequence of our insistence upon the equivalence principle in a narrow spacetime region at the cost of conformal invariance. In Section 4, we summarize the definite conclusions based on the above discussion. It is seen that the solution to these two crucial problems, \emph{viz.,} of singularity and energy calls for the exclusion of the zero space and zero time(epoch of singularity) from the action of GTR which it itself does not command.

\section{Singularity as a Non-Machian\\ Feature}

Since there exist at least 10 different versions of the Mach's Principle \cite{f1,m1} , first we describe our version of the same adopted
in the present discussion. To fix ideas, we take principle as  ``no background''(A) $\Leftrightarrow$ ``no matter''(B). While \begin{eqnarray}A\rightarrow B\label{n1} \end{eqnarray}
is commonly understood as the primary requirement of the Mach's principle, the complementary interaction in form of \begin{eqnarray}B\rightarrow A \label{n2}\end{eqnarray}
is provided by an inherent symmetry underlying the causal connections.However, none of these two conditions is true in the GTR or the Newtonian framework.
From the variation of the Einstein-Hilbert action
\begin{eqnarray} S &=&\frac{1}{16\pi G}\int (R+2\lambda)\sqrt{-g}{d^4}{x}+\int L_{phys}\sqrt{-g}d^4 x\label{n3}\end{eqnarray}
with respect to the metric we obtain the field equations
\begin{eqnarray}R_{ik}-\frac{1}{2}g_{i k}R+\lambda g_{ik}&=&- 8\pi G T_{ik}.\label{n4}\end{eqnarray}

In the absence of material background in this theory, we still have a spacetime structure given by
\begin{eqnarray}R_{ik}=0\label{n5}\end{eqnarray}
against which any particle may be introduced along its world-line. Such a particle has its own dynamics not supported by any material background. It is in contradiction with (\ref{n1}). Secondly, the masses of the system of particles do not evolve together in a mutual inter-particle or field response which shows the violation of (\ref{n2}). Under the conformal transformation on
(\ref{n3}) from manifold $M \rightarrow \tilde{M}$
\begin{eqnarray}\tilde{g_{ik}}=\Omega^2(X^i){g_{ik}}\label{n6}\end{eqnarray}
where ${g_{ik}}$ and $\tilde{g_{ik}}$are the metrics in the manifolds $M$ and $\tilde{M}$ respectively, $\Omega(X^i)$ is a twice differentiable function of the coordinates $X^i$ and $0<\Omega<\infty$, we have the new scalar curvature as
\begin{eqnarray}\tilde{R}=\Omega^{-2}(R-{6}\Omega^{-1} \Box \Omega)\label{n7} \end{eqnarray}
with determined with respect to the variations on the metric. The GTR thus lacks conformal invariance.

Now if in(\ref{n4}) we introduce a scale change on the mass functions

\begin{eqnarray}\Omega (X^i)=\frac {M(X^i)}{\tilde{m}}\label{n8}\end{eqnarray}
we find that these remain unaffected because of the constancy of masses in Einstein frame and so the violation of Mach principle. Since singularity is unavoidable in GTR as shown by the Penrose-Hawking theorems \cite{b1} it can exist only with $\tilde{m}\rightarrow\infty $(or $\tilde{R}\rightarrow\infty$) consequent to the violation of the conformal condition \begin{eqnarray}\Omega\neq0.\label{n9}\end{eqnarray}
Clearly we have infinite energy with no background (which already vanishes under the effective condition on the scale change) in violation of (\ref{n1})and (\ref{n2}). In this way one can actually have infinitely many possible frames corresponding to their unique mass functions each violating (\ref{n9})and containing a singularity.

In Einstein frame, non-zero mass functions may be achieved under the above ``forced'' conformal transformation from zero mass hypersurfaces by breaking (\ref{n9}). Clearly, two indications appear from this observation, as has been previously pointed out \cite{m2}. First, singularity is a non-Machian feature in GTR  with  mass functions blowing up $\tilde{m}=\infty$, and the Compton length scale turning into a singularity. Secondly, it violates the conformal condition (\ref{n9}).

As a conceptual  alternative, a  more general approach that uses  both (\ref{n1}) and (\ref{n2}) is provided by the Hoyle-Narlikar inter-particle interaction theory (at the base of QSSC \cite{g2,g3}) with its equations given by
\begin{eqnarray}R_{ik}-\frac{1}{2}g_{i k}R+\lambda g_{ik} & = & -\frac{6}{M^2}[T_{ik}-\frac{1}{6}(g_{ik} M^2-M_{;ik}^2)\nonumber \\
& & -(M,_{i}M,_{k}-\frac{1}{2}g_{ik}g^{pq}M,_{p}M,_{q})].\label{n10}\end{eqnarray}
where $M,_{i}\equiv \frac{\partial M}{\partial X^i}$ are the derivatives of the mass functions M and other symbols have their usual meaning.

 It is can be readily seen from (\ref{n10}) that if the scalar mass functions  remain fixed under the scale change (\ref{n8}) these reduce to the non-Machian Einstein equations (\ref{n4}).
\vskip.15in
\noindent
 We emphasize the following points...
\begin{enumerate}
\item[(i)] Some authors have mentioned that the singularity results from the occurrence of the zero-mass hypersurfaces in the gravity equations leading to unphysical effects \cite{n1}. It is obvious, however, that a conformal transformation without violating (\ref{n8})can still be invoked for zero mass hypersurfaces and singularity may be averted. This will be ``empty to empty'' transformation. It is like Milne's empty but singularity-free model with curvature parameter given as $k=-1$. But we find it non-trivial and will discuss in Section 3.

\item[(ii)] The above facts have been  brought about not for a comparison between GTR and QSSC but just in order to discuss the challenging issues of singularity(and energy in Section 3) in GTR and the fundamental requirement of Machian connections and conformal invariance to avert them. It appears that the conformal invariance is a necessary condition to a singularity-free theory, although the converse is not true as is evident from several singularity-free solutions given in  the spatially inhomogeneous cosmological models \cite{d1,e1}. Even though,these solutions represent the complete causal curves with well defined cylindrical symmetry, it is found on closer inspection that they do not satisfy the assumption of compact trapped surfaces in the Penrose-Hawking theorems for the exact perfect fluid given by $p=\frac{1}{3}\rho$. Several authors have attempted to avoid singularity in the spherical models too where it becomes possible if the instrumental role of shear in collapse in the Raychaudhury equations is surpassed by the counter-acceleration\cite{o1}.
\end{enumerate}
 It may be noted that a common condition frequently used in the family of  cylindrical or spherical models is of inhomogeneity of spacetime(perturbed Friedman-Roberson-Walker metric)\cite{e1}. Clearly, such condition can be used without invoking the basic equations of the theory, i.e., GTR in the present case and is independent of the action which is already not scale invariant as discussed above and actually does not require by itself the removal of singular epoch, since it is determined over all spatial volume between no preferred choice of temporal limits, including $t=0$ epoch. Therefore it may not be justified to again put the \emph{extra} conditions on this action to exclude singular epoch. Similarly, if we drop the assumption of the compact trapped  surfaces which was motivated by the argument that the energy density needed to thermalize the Cosmic Microwave Background Radiation (CMBR) is sufficient enough to converge the past geodesic congruence,we have no singularity in GTR \cite{c1}. This means that zero mass hypersurfaces might exist in some conformal frames but not in all. With either approach, we feel that any avoidance of singularity must descend directly from the dynamical theory and not from the choice of perturbed metric or the subjective dropping of conditions in the singularity theorems.

\section{Conformal Invariance and the Non-Conservation of Energy}

The definition of energy in GTR is another challenging problem which has attracted the attention of many authors \cite{h1}. It is understood that in presence of the gravitational fields, vanishing of the four-divergence of matter energy-momentum tensor alone, \emph{i.e.}
\begin{eqnarray}T_{i;k}^k=0\label{n11}\end{eqnarray}
carries no physical meaning for the conservation of energy\cite{i1}. Thus to include the gravity another pseudo-tensor $t_{ik}$ \cite{p1}(or tensor as in the field formulations by several authors\cite{l1,l2,l3} )is invoked as

\begin{eqnarray}T_{ik}\rightarrow(-g)(T_{ik}+t_{ik}).\label{n12}\end{eqnarray}

Since $t_{ik}$ in (\ref{n12}) is not gauge invariant, it cannot define energy without additional constraints of asymptotic flatness, as independently used in ADM (Arnowitt-Deser-Misner) formalism \cite{j1} or static metric(Komar masses \cite{k1}). While one has to include the functional dependence of the deviations of metric from the asymptotic spacetime to retrieve the ADM energy, the theory has no way to link them with the global structure. It does not work at least on three counts. One, in the actual universe with the observed large scale structures,asymptotic flatness is ruled out. Two, in absence of any Machian connections (\ref{n1}) or (\ref{n2}), except for very weak gravity waves, no causal communication can be established between any masses in the large scale regions and the local spacetime. It is because an arbitrary tube around a particle world-line cutting through the spacetime never has zero gravity as $r\rightarrow\infty$ in a homogeneous matter distribution without potential causal links. Here the global structure of light cones is not preserved.Three, ADM energy conservation, as also the Komar masses, require time independent asymptotic metric whereas larger distances go into higher redshift structure evolution with strong metric perturbations and approaching the singular epoch $t=0$. Time translational symmetry is not obeyed here and by Noether's theorem the conservation of energy breaks down.

Against this background, we argue that this inadequacy in localization of energy (or the corresponding attempts to recover a global conservation of energy with no apparent causal links), which is in fact an immediate consequence of the equivalence principle, can only be avoided by excluding from (\ref{n3}) the zero (so-called infinitesimally small)
spacetime volume around the spacetime point where we want to localize the energy. But as discussed in Section 2 we include  entire spatial volume through arbitrary time bounds in (\ref{n3}). Thus, here the similar problem crops up with the spacetime point with a zero volume, though not excluded by GTR yet manifesting itself in form the non-localizability of energy. The equivalence principle pins down to a spacetime point where the energy vanishes. This is not surprising to us because this is precisely the matching situation that led to a $t=0$ singularity in Section 2 by breaking the conformal condition (\ref{n9}). Thus the theory lays no \emph{a priori} restrictions on the initial choice of the any spacetime point for calculations of local gravitational energy, but we find  it vanishes there. To restore the local energy, it is inevitable to exclude that spacetime point from (\ref{n3})  and thus to sacrifice the equivalence principle. Or else, if we retain the equivalence principle then conformal invariance (in general,any kind of conformal transformation for $\Omega=0$) must be abandoned, as is the case in GTR. Since quantum field theories are already conformally invariant,
 we expect that the two requirements, \emph{viz.} the equivalence principle and the conformal invariance are mutually inconsistent. This may be one posssible reason for difficulty in the compatibility of GTR with quantum formulation of gravitation.

In a gravitational theory,such as (\ref{n3}) the effective tensor of matter must include the \emph{in built} gravitational field contributions to energy and momentum which must not arise from outside terms like $t_{ik}$. Since it is generated by $L_{phys}$ as $T_{ik}\equiv 2\left(\frac{\delta L_{phys}}{\delta g_{ik}}\right)$ that is a source of gravity in the action (\ref{n3}), it is difficult to think of ``gravity-free'' $T_{ik}$ of matter fields alone. On the other hand,in a singularity-free, conformally invariant QSSC  \cite{g1,g2,g4} with equations (a form of (\ref{n10}))with the creation fields C)
\begin{eqnarray}R_{ik}-\frac{1}{2}g_{ik}R+\lambda g_{ik}=-8\pi G[T_{ik}-f(C_{,i}C_{,k}-\frac{1}{4}g_{ik}C^{,l}C_{,l}]\label{n13}\end{eqnarray}
we have the net divergence of the right hand side being zero, giving $T^{ik}_{;k}=fC^{i}C^{k}_{;k}$ keeping the energy conservation intact both in creative mode$(T^{ik}_{;k}\neq0)$ and the non-creative mode $(T^{ik}_{;k}=0).$ Here, we do not need any terms except those,(such as matter, electromagnetic radiation or C-field) that incorporate gravity to realize the conservation of energy and momentum. Apart from being Machian and conformally invariant to keep the global light-cone structure invariant, it has clear advantages over the above-mentioned efforts in GTR to retrieve energy  by making untenable assumptions of asymptotic flatness or static metric. Since this theory is not based on the equivalence principle, there does not exist any conflict with the requirement of conformal invariance in view of our above arguments.

Here we may also recall a scenario like Milne's $k=-1$ empty universe mentioned in Section 2. Avoidance of singularity and global conservation of energy both may be satisfied in an otherwise empty universe with net energy density $\rho=0$, where mass degeneracy breaks into two symmetrical states, positive $\rho_{+}$ and negative $\rho_{-}$. The second component $\rho_{-}$ with positive equation of state violates the weak energy condition but due to its negative pressure must drive the large scale cosmic acceleration like the  quintessence scalar field while the $\rho_{+}$ acts in matter creation.

\section{Summary}

To sum up, we have argued in this paper that the avoidance of singularity or conservation of energy must descend from the dynamical theory itself and not from the subjective assumptions on spacetime metric. We have attempted to establish the connected  arguments in construction of groundwork for a viable theory of gravity. Since we know GTR includes a singularity and makes any physical laws there go awry, we showed that it results from the lack of fulfilment of Machian principle in form of (\ref{n1}) and (\ref{n2}) and together these do not allow the dynamical theory to be conformally invariant.

In Section 3 we have shown that the equivalence principle in form of the non-localization  of energy in GTR (leading to the non-conservation of global energy) is incompatible with the requirement of conformal invariance. Together with a similar problem cropping up in form of singularity in Section 2 this exhibits a  handicap on the basic action(\ref{n3}) where we have to remove the ``aching'' points from it. In purely scientific spirit, our clinical analysis brings in to focus the basic requirements of Machian connections and conformal invariance with the example of an alternative theory of QSSC models which is free from these problems in its basic structure.\\
\vskip.2in
\noindent
\textbf{\large Acknowledgments }\\

\noindent
The author is thankful to P. Creminelli for helpful discussions and to  The Abdus Salam International Centre for Theoretical  Physics, Trieste, Italy for its kind facilities where this work was completed under the federation arrangement.

\end{document}